\documentstyle[twocolumn,floats,epsf,ifthen,aps,prb]{revtex}

\begin{document}


\draft

\title{Spectrum of qubit oscillations from Bloch equations} 

\author{Rusko Ruskov\footnote{On leave of absence from
Institute of Nuclear Research and Nuclear Energy, Sofia BG-1784,
Bulgaria} and Alexander N. Korotkov\footnote{Electronic mail:
korotkov@ee.ucr.edu}}
\address{
Department of Electrical Engineering, University of California,
Riverside, CA 92521. }
\date{\today}

\maketitle

\begin{abstract}
We have developed a formalism suitable for calculation of the output 
spectrum of a detector continuously measuring quantum coherent oscillations 
in a solid-state qubit, starting from microscopic Bloch equations.  
  The results coincide with that obtained using Bayesian and 
master equation approaches. The previous results are generalized 
to the cases of arbitrary detector response and 
finite detector temperature.

\end{abstract}

\narrowtext

\vspace{0.6cm}

\section{ Introduction}

        Quantum coherent (Rabi) oscillations in a two-level system 
represent a simple and fundamental example of a nontrivial quantum 
behavior.\cite{Rabi} Recently increased interest in this subject 
is obviously related to the use of two-level systems (qubits) as
building blocks of a prospective quantum computer.\cite{Bennett}
The emphasis has naturally shifted from traditional observations 
of Rabi oscillations in ensembles of two-level atoms to the studies
of single qubits. Concentrating in this paper on solid-state qubits, 
let us mention recent demonstrations of single-qubit quantum 
coherent oscillations in both time\cite{Nakamura} and 
frequency\cite{Mooij,Lukens} domains. 
(Rabi oscillations in individual quantum dots have been also demonstrated
by traditional optical means;\cite{Stievater} however, we will discuss only 
solid-state qubits with electronic readout.) 

\begin{figure} 
\centerline{
\epsfxsize=2.0in 
\epsfbox{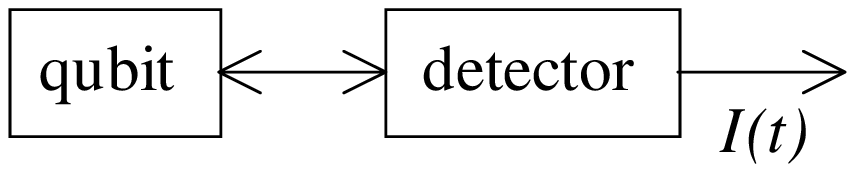}
} 
\vspace{0.4cm} 
\caption{Schematic of a single solid-state qubit continuously measured 
by a detector.
 }
\label{Fig1}\end{figure}

        Even though experiments of Refs.\ \cite{Nakamura,Mooij,Lukens} 
have been done with single qubits, their results are essentially 
ensemble-averaged, and as a consequence, the problem of a quantum state 
collapse due to measurement is not very important. Another possible
experimental setup (for which the collapse is of the major importance) 
is a {\it continuous} monitoring of single-qubit quantum 
oscillations (Fig.\ \ref{Fig1}). Such setup is similar to that used
in experiments \cite{Buks,Sprinzak}, with the difference that the detector
output signal $I(t)$ was actually not studied experimentally. 

        The basic questions in prospective experiments of the type shown 
in Fig.\ \ref{Fig1} are the following: 1) what is the effect of continuous
quantum measurement on the qubit evolution, 2) how the detector output $I(t)$
looks like, and 3) what is the relation between detector output and
qubit evolution? Some answers to these questions have been obtained recently 
(see, e.g., Ref.\ \cite{Kor-rev} and references therein); however, the subject
is still active and quite controversial. (The main reason of controversy is 
the necessity to go beyond the Schr\"odinger equation to describe continuous 
collapse due to measurement.) 

     In this paper we will consider a relatively simple question: 
what is the spectral density $S_I(\omega )$ of the detector output $I(t)$
and how high is the spectral peak corresponding to quantum oscillations of
the qubit state? 
(Notice that we assume detector output $I(t)$ to be a classical magnitude,
so $S_I(\omega )$ does not depend on a particular method of further signal
processing.) 
This question has been addressed already in a number of papers (see, e.g., 
Refs.\ 
\cite{Kor-osc,Kor-Av,Av-book,Goan-osc,Fazio,Av-osc2,Mozyrsky,Hackenbroich,%
Makhlin-PRL}) 
using various techniques. 

   Spectral density $S_I(\omega )$ has been calculated using Bayesian 
formalism
\cite{Kor-rev,Kor-99} in Ref.\ \cite{Kor-osc} for the case of a weakly 
responding (linear) detector. In particular, it has been shown that the 
spectral
peak $S_I(\Omega )$ at the frequency $\Omega$ of quantum oscillations cannot 
be higher than $4S_0$ where $S_0$ is the noise pedestal due to intrinsic
detector noise. 
     It has been also shown that the results of the Bayesian 
formalism for $S_I(\omega )$ exactly coincide with the results 
\cite{Kor-osc,Kor-Av} obtained using the standard master equation formalism.
In spite of the same results, the interpretations are quite different 
since the Bayesian formalism describes individual monitoring of quantum 
evolution in time and treats $I(t)$ as a classical measurement result, 
while the master 
equation formalism can describe only ensemble-averaged magnitudes and 
in some sense should treat $I(t)$ as a quantum operator. 
(The results of two formalisms coincide because $S_I(\omega )$ is essentially
an average quantity. In more general situations, for example, for a quantum
feedback analysis\cite{Kor-rev,Ruskov} the results are not the same since 
the master equation formalism fails.) 

        The results for $S_I(\omega )$ have been confirmed in Ref.\ 
\cite{Av-book} using a somewhat different approach based on the general 
theory of linear detectors. In Ref.\ \cite{Goan-osc} the results have been
confirmed using the approach of quantum trajectories\cite{Milburn-93,Goan} 
adopted from quantum optics (this approach is similar to the Bayesian
formalism). It has been also shown \cite{Goan-osc} that the same formulas
for $S_I(\omega )$ remain valid even when the condition of weakly responding 
detector is not satisfied (the detector response $\Delta I$ to the 
change of the qubit state is comparable to the average current $I_0$). 

        A different result for $S_I(\omega )$ has been obtained in Ref.\
\cite{Fazio} (the studied system is slightly different from the type shown
in Fig.\ \ref{Fig1}; however, it is essentially similar). In particular,
the calculated ratio $[S_I(\Omega )-S_0]/S_0$ can be arbitrary large
(not limited by 4). In our opinion, this is because the result of Ref.\ 
\cite{Fazio} includes the contribution from zero-point oscillations 
which are not measurable in a straightforward way. In principle, zero-point 
oscillations can be sensed in an experiment; however, it would necessarily 
require quantum (non-classical) interaction between the detector 
[which outputs $I(t)$] and the next stage. In other words, it would require
measurement of $S_I(\omega )$ without ever measuring $I(t)$. 
A simple example of such setup is the absorption/emission of photons 
at resonant frequency $\Omega$. A more sophisticated example of such 
measurement 
of qubit oscillations has been considered in Ref.\ \cite{Av-osc2}. 
The idea is to use rotating measurement basis, in which there are essentially
no oscillations, but rather jumps between two stationary states due to
external noise. Formally shifting the zero-frequency spectrum of such
continuous measurement to the frequency $\Omega$, one can obtain arbitrary
high spectral peak. (There is no restriction on the ratio 
$S_I(\omega =0)/S_0$ 
in a strong dephasing case.\cite{Kor-osc,Kor-Av,Av-book,Goan-osc})
Such setups, however, are not the subject of the present paper since
we limit ourselves to the straightforward case of Fig.\ \ref{Fig1}
with $I(t)$ being usual classical signal which can be amplified further 
by any good amplifier. Also, we do not consider here the 
quantum feedback setups, which can provide arbitrary high spectral peak
of the detector current at the oscillation frequency $\Omega $.\cite{Ruskov}

       The main result $[S_I(\Omega )-S_0]/S_0 \leq 4$ for a simple setup 
seems to contradict the experimental results of 
Ref.\ \cite{Manassen} which claim the measurement 
of $S_I(\omega )$ from a single spin precession in an STM-based setup 
(a significantly different, but still an analogous experiment). 
In a similar recent experiment \cite{Durkan} the maximum observed 
peak-to-pedestal ratio for a measurement of a single spin precession 
was a little less than 4.\cite{Durkan-2} 
We cannot explain the disagreement between the theory and experiment
of Ref.\ \cite{Manassen}; however, we note that in the recent theoretical 
paper \cite{Mozyrsky} which considers a somewhat similar setup, the 
possibility of a relatively high spectral peak has not been confirmed. 
The spectral peak of $S_I(\omega )$ due to qubit oscillations has been
also considered theoretically in Ref.\ \cite{Hackenbroich}; however, 
the peak magnitude has not been calculated. 

        Let us finally mention one more theoretical approach (developed by 
S.\ A.\ Gurvitz) to the analysis of 
detector output $I(t)$ based on Bloch equations which describe the evolution
of the coupled ``qubit+detector'' density matrix.\cite{Gurvitz} 
The advantage of this approach is the straightforward microscopic
derivation of Bloch equations from the Schr\"odinger equation, 
while in the approaches mentioned above (Bayesian, master equation, 
quantum trajectory) the collapse ansatz is either explicitly or implicitly
used. The Bloch equation approach has been used to calculate some statistical
characteristics of $I(t)$ in Ref.\ \cite{Makhlin-PRL}; however, the spectral
density $S_I(\omega )$ has been obtained only at the frequencies lower than
the ensemble dephasing rate and so the spectral peak due to quantum 
oscillations has been out of scope of Ref.\ \cite{Makhlin-PRL}.

        In this paper we show how the Bloch equations can be used to calculate
the detector output spectral density $S_I(\omega )$ for a particular 
measurement setup (Fig.\ \ref{Fig2}). The case of an arbitrary qubit 
coupling with detector and finite detector temperature is considered. 
 We prove that the results for $S_I(\omega )$ coincide 
with that obtained previously (in a narrower validity range) 
by the master equation and Bayesian 
approaches. The Bayesian results are generalized to the case of 
arbitrary response factor and finite detector temperature; 
it is shown that the equivalence of results of 
the three approaches still holds in this case.

\section{ The system and Bloch equations} 

\begin{figure} 
\centerline{
\epsfxsize=2.0in 
\epsfbox{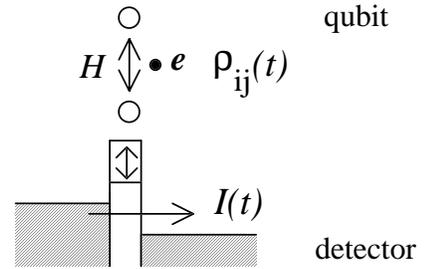}
} 
\vspace{0.4cm} 
\caption{Small transparency QPC (tunnel junction) as a qubit state detector. 
The barrier height depends on the electron position in the double-dot system.
 }
\label{Fig2}\end{figure}

        We consider the system (Fig.\ \ref{Fig2}) introduced in Ref.\ 
\cite{Gurvitz} and studied extensively after that. 
\cite{Buks,Sprinzak,Kor-rev,Kor-osc,Kor-Av,Av-book,Goan-osc,Hackenbroich,%
Kor-99,Ruskov,Goan} 
The qubit is represented by a single electron in a double quantum dot. 
The detector is a quantum point contact (QPC) whose barrier height depends 
on the electron position, so the current $I$ through QPC measures the qubit
state in the basis of localized states $|1\rangle$ and $|2\rangle$. 
We will limit ourselves by the case of small transparency QPC 
which is equivalent to a simple tunnel junction.
The Hamiltonian of the system,
        \begin{equation}
{\cal H} = {\cal H}_{QB}+{\cal H}_{DET} +{\cal H}_{INT},
        \label{Hamiltonian}\end{equation} 
describes the qubit, detector, and their interaction: 
        \begin{eqnarray}
&& {\cal H}_{QB} = \frac{\varepsilon}{2}\, (c_1^\dagger  c_1 - c_2^\dagger 
c_2)
     + H \, (c_1^\dagger c_2 +c_2^\dagger c_1) , 
        \label{Hqb}
        \\ 
&& {\cal H}_{DET} = \sum_l E_l a_l^\dagger a_l +\sum_r E_r a_r^\dagger a_r 
        \nonumber\\
 &&\hspace{1.3cm} 
+\sum_{l,r} M \, a_l^\dagger a_r+ \mbox{H.c.}\, ,  
        \label{Hdet}\\ 
&& {\cal H}_{INT}= \sum_{l,r} \frac{\Delta M}{2} \, 
(c_1^\dagger c_1 - c_2^\dagger c_2) \, 
a_l^\dagger a_r + \mbox{H.c.} 
        \label{Hint}\end{eqnarray}
(for simplicity $M$ and $\Delta M$ are assumed to be real and 
energy-independent). 
        The average detector currents corresponding to the qubit states 
$|1\rangle$ and $|2\rangle$ are equal to 
$I_1= 2\pi (M+\Delta M/2)^2 \rho_l\rho_r e^2V/\hbar$  and
$I_2=2\pi (M-\Delta M/2)^2 \rho_l\rho_r e^2V/\hbar$, correspondingly 
($V$ is the voltage across the tunnel junction, $e$ is the electron charge,
and $\rho_{l,r}$ are 
the densities of states in the electrodes), while the corresponding
detector noises have white spectrum and are given by the Schottky formula:
        \begin{equation}
S_{1,2}=2eI_{1,2} .
        \end{equation} 
Note that the detector voltage $V$ is assumed to be large enough,
so that the typical quantum noise frequency $eV/\hbar$ is much higher 
than all relevant frequencies.\cite{eV}

In the following we will distinguish the weakly-responding limit,
$|\Delta I|\ll I_0$, where $\Delta I\equiv I_1-I_2$ is the detector 
response  
and $I_0\equiv (I_1+I_2)/2$, and the finite response case, 
$|\Delta I| \sim I_0$. Notice that the word ``coupling'' is reserved for a
different combination of parameters: ${\cal C}\equiv \hbar (\Delta I)^2/S_0H$
[here $S_0\equiv (S_1+S_2)/2$], which affects the quality factor of quantum
oscillations of the qubit. The frequency of unperturbed oscillations 
(without detector) is
equal to $\Omega \equiv \sqrt{4H^2+\varepsilon^2}/\hbar$, where $H$ is the
qubit tunneling matrix element (assumed to be real) and $\varepsilon$ 
is the qubit energy asymmetry.

        Our starting point is the Bloch equations\cite{Gurvitz} describing  
the {\it ensemble averaged} 
evolution of the density matrix $\rho_{ij}^n$ in which the subscripts 
($i,j=1,2$) label the qubit state while $n$ is the number of electrons 
passed through the detector (only diagonal in $n$ matrix elements are 
considered because the nondiagonal elements decay very fast). 
For our system the Bloch equations are the following: \cite{Gurvitz,Kor-rev}
        \begin{eqnarray}
 \dot\rho_{11}^{\,n} = && - \frac{I_1^+ +I_1^-}{e} \, \rho_{11}^n 
        + \frac{I_1^+}{e}\, \rho_{11}^{n-1} 
        + \frac{I_1^-}{e}\, \rho_{11}^{n+1} 
                \nonumber \\
&&  {}    -2\, \frac{H}{\hbar} \, \mbox{Im}\, \rho_{12}^n \, ,
        \label{Bloch1}\\ 
 \dot\rho_{22}^{\,n} = && - \frac{I_2^+ +I_2^-}{e} \, \rho_{22}^n 
        + \frac{I_2^+}{e}\, \rho_{22}^{n-1}  
        + \frac{I_2^-}{e}\, \rho_{22}^{n+1}  
                \nonumber \\ 
&&  {}    + 2\, \frac{H}{\hbar} \, \mbox{Im}\, \rho_{12}^n  \, , 
        \label{Bloch2}\\ 
 {\dot\rho}_{12}^{\, n} = && 
        -\frac{I_1^+ + I_1^- +I_2^+ +I_2^-}{2e} \, \rho_{12}^n    
	+ \frac{\sqrt{I_1^+I_2^+}}{e} \, \rho_{12}^{n-1}
        \nonumber \\
&&   {}   
        + \frac{\sqrt{I_1^-I_2^-}}{e} \, \rho_{12}^{n+1} \, 
        + {\rm i} \, \frac{\varepsilon}{\hbar} \,\rho_{12}^n+ 
        {\rm i} \,\frac{H}{\hbar } \,(\rho_{11}^n-\rho_{22}^n) .    
        \label{Bloch3}\end{eqnarray}
Here 
\begin{equation}
I_i^+ = \frac{I_i}{1-\exp (-eV/T)}\, , \,\,\,
I_i^- = I_i^+ \exp (-eV/T) 
\end{equation}
are the partial currents in two directions ($I_i=I_i^+-I_i^-$) and
$T$ is the detector electron temperature. Notice that this temperature is 
different from what was considered in Refs.\ \cite{Kor-osc,Kor-Av,Av-book}. 
In those papers 
the effect of nonzero temperature of a passive environment coupled to the 
qubit was studied, so the important parameter was $T/\hbar\Omega$. 
In this paper (similar to Ref.\ \cite{Goan}) we consider the effective 
detector temperature and the important parameter is $T/eV$, while a finite 
temperature $T$ always implies $T\gg \hbar\Omega$. (The phonon temperature
in the vicinity of the qubit may still be low, since $T$ is only the 
electron temperature.)
The density matrix $\rho_{ij}^n$ obeys the natural normalization condition 
$\sum_n (\rho_{11}^n+\rho_{22}^n)=1$. 

        Notice that tracing Bloch equations (\ref{Bloch1})--(\ref{Bloch3})
over the detector degree of freedom $n$, one obtains a conventional master
equation for the qubit:
\begin{eqnarray}
&&\dot{\rho}_{11}=    -2 \, \frac{H}{\hbar} \,\mbox{Im} 
\rho_{12}, \,\,\,\, \rho_{11}+\rho_{22}=1, 
        \label{conv1}\\
&& {\dot\rho}_{12}=  {\rm i} \, \frac{\varepsilon}{\hbar}\, \rho_{12}+ 
        {\rm i} \, \frac{H}{\hbar }\, 
(\rho_{11}-\rho_{22})
-\Gamma \rho_{12},  
        \label{conv2} 
\end{eqnarray}
        where $\rho_{ij}=\sum_n\rho_{ij}^n$ and 
the ensemble decoherence rate $\Gamma$ is equal to \cite{Gurvitz,Goan} 
\begin{eqnarray}
\Gamma =&& \frac{\left(\sqrt{I^+_1}-\sqrt{I^+_2}\right)^2}{2e} + 
         \frac{\left(\sqrt{I^-_1}-\sqrt{I^-_2}\right)^2}{2e} 
\nonumber\\
 && = \frac{(\sqrt{I_1}-\sqrt{I_2})^2}{2e}\,\coth (eV/2T)\, .
\label{Gamma} 
\end{eqnarray}
(The decoherence rate $\gamma$ for a single system without ensemble
averaging is different -- see Section V.)

        \section{Spectral density via MacDonald's formula}

        The Bloch equations couple the qubit evolution and the 
number $n$ of electrons passed through the detector. So, to calculate
the spectral density $S_I(\omega )$ of the detector current, we need 
to express $S_I(\omega )$ in terms of $n$. This can be easily done
for the {\it classical} random process $I(t)$ using the MacDonald's 
formula\cite{MacDonald}
\begin{equation}
S_I(\omega)=2\, \omega\, \int_0^{\infty} \,
\frac{d \langle Q^2(\tau )\rangle}{d \tau }\, \sin (\omega \tau )\, d\tau 
\label{mcd}\, ,
\end{equation}
        where $\langle Q^2(\tau )\rangle = \langle \left( \int_t^{t+\tau}
I(t')dt' -\langle I\rangle \tau \right)^2 \rangle$ and averaging is over 
time $t$ (MacDonald's formula have been also used in Ref.\ \cite{Mozyrsky}). 
In our case the average current is equal to $I_0$ (see below) for 
a non-zero qubit tunneling $H$, so 
        \begin{equation}
\langle Q^2(\tau ) \rangle = e^2 \langle n^2(\tau ) \rangle - (I_0 \tau)^2 ,
        \label{mcd2}
        \end{equation}
 where $\langle n^2(\tau )\rangle$ is the average square of the number of
electrons passed through the detector during time interval $\tau$.

        To calculate $\langle n^2(\tau )\rangle$ we can use the Bloch 
equations
and the obvious relation 
        \begin{equation}
\langle n^2(\tau )\rangle = \sum_n n^2 [\rho_{11}^n(\tau )+
        \rho_{22}^n(\tau )]. 
        \label{n2}\end{equation}
However, the situation is not too simple because the left hand side
contains the averaging over time while the right hand side is essentially
ensemble averaging which depends on the initial condition $\rho_{ij}^n (0)$.
Quite naturally we should assume $n=0$ at $\tau =0$, so that 
$\rho_{ij}^n (0)=\delta_{n0}\rho_{ij}(0)$, but the question about the choice
of $\rho_{ij}(0)$ remains unclear because the qubit state actually 
oscillates in time (for a nonzero $H$, which case we always assume below). 
A natural choice is to use the stationary value: 
$\rho_{ij,st}=\lim_{t\rightarrow \infty}\sum_n [\rho_{ij}^n(t) +
\rho_{ij}^n(t)]$, and it is possible to prove that this choice is really 
correct in the following way.

        As we know from the Bayesian formalism,\cite{Kor-rev} we can 
monitor the oscillating
evolution of the qubit density matrix $\rho_{ij}(t)$ in an individual 
realization of the experiment using the detector output $I(t)$. 
This at least means that $\rho_{ij}(t)$ exists (even though it cannot be 
obtained using Bloch equations because they imply ensemble averaging).
So, the correct procedure of calculating $\langle n^2(\tau )\rangle$ 
would be the
following. The right hand side of Eq.\ (\ref{n2}) should be calculated
for various initial values $\rho_{ij}(\tau =0)$ corresponding to values
$\rho_{ij}(t)$ in a sufficiently long realization of a process,
 and then the result should 
be averaged over the time $t$ [i.e. weighted proportionally to  the 
occurrence 
frequency of various $\rho_{ij}$]. Now it is very important that the
Bloch equations (\ref{Bloch1})--(\ref{Bloch3}) are linear in respect to
the initial condition. This means that instead of averaging the result for 
$n^2(\tau )$ over initial condition $\rho_{ij}(\tau =0)$, 
we can use the initial condition which is itself 
the value averaged over time, i.e. stationary value $\rho_{ij,st}$ 
discussed above (of course, 
we implicitly use the process ergodicity). This ends the proof.

        {\it Thus}, to calculate $S_I(\omega )$ we should solve the Bloch 
equations starting from the stationary initial condition 
$\rho_{ij}^n(0)=\delta_{n0}\rho_{ij,st}$,
then calculate $\langle n^2(\tau )\rangle$ using Eq.\ (\ref{n2}),
and then use MacDonald's formula (\ref{mcd}) to obtain $S_I(\omega )$.
Notice that the stationary state $\rho_{ij,st}$ can be easily obtained
from the master equations (\ref{conv1})--(\ref{conv2}) and the condition
$\dot{\rho}_{ij}=0$, that gives (at $H\neq 0$) 
\begin{equation} 
        \rho_{11,st}=\rho_{22,st}=1/2, \,\,\, \rho_{12,st}=0.
\end{equation} 
(The stationary state would be different if the qubit had an extra 
coupling to a passive environment;\cite{Kor-osc,Kor-Av} however, 
we do not consider such case.)

        One can use this method to calculate $S_I(\omega )$ in 
a straightforward way
(we have done it numerically); however, it is better to use an analytical
simplification calculating directly $d\langle n^2(\tau )\rangle /d\tau
=\sum_n n^2[\dot{\rho}_{11}^{\,n}(\tau )+\dot{\rho}_{22}^{\,n}(\tau )]$.
Using Eqs.\ (\ref{Bloch1}) and (\ref{Bloch2}) and shifting summation over
$n$ in terms containing $\rho_{ii}^{n\pm 1}$, one gets the equation
\begin{eqnarray}
\frac{d\langle n^2(\tau )\rangle }{d\tau} = &&
\frac{I_0}{e}\left( 2\langle n(\tau )\rangle+\coth \frac{eV}{2T}\right) 
        \nonumber\\
+ && \frac{\Delta I}{e} \left( {\cal A}(\tau ) + \frac{z(\tau )}{2} 
\, \coth \frac{eV}{2T}\right), 
        \label{dn2dt}\end{eqnarray}
where 
\begin{eqnarray}
&& {\cal A}(\tau ) \equiv  \sum_n n\, 
        [\rho_{11}^n(\tau )-\rho_{22}^n(\tau )]  ,
        \label{m-def} \\ 
&& \langle n(\tau ) \rangle \, \equiv \sum_n n\, 
        [\rho_{11}^n(\tau )+\rho_{22}^n(\tau )], 
		\\
&& z(\tau )\equiv \sum_n [\rho_{11}^n(\tau )-\rho_{22}^n(\tau )].
\end{eqnarray}
Notice that $z(\tau )=0$ since the evolution starts from the stationary 
state, $\rho_{ij}^n(0)=\delta_{n0}\rho_{ij,st}$, so the corresponding term 
in Eq.\ (\ref{dn2dt}) vanishes. 

To calculate $\langle n(\tau )\rangle$, we again use Eqs.\ 
(\ref{Bloch1}) and (\ref{Bloch2}), shift summation over
$n$, and obtain the equation 
\begin{equation}
        d\langle n(\tau )\rangle /d\tau = I_0 +z(\tau )\Delta I/2. 
\end{equation}
Since the last term vanishes because of $z(\tau )=0$ and since 
$\langle n(0)\rangle =0$, we obtain a simple result 
$\langle n(\tau )\rangle =I_0\tau$. 
In particular, this means that the average detector current is equal to
$I_0$ (we have used this result above). 

        One can see that the term $2I_0\langle n(\tau )\rangle /e=
 2I_0^{\,2}\tau$ from Eq.\ 
(\ref{dn2dt}) exactly cancels the contribution from the derivative of the 
last term of Eq.\ (\ref{mcd2}). The term $(I_0/e)\coth (eV/2T)$ from 
Eq.\ (\ref{dn2dt}) after being plugged into MacDonald's formula
(\ref{mcd}) gives the constant noise pedestal $2eI_0\coth (eV/2T)$ 
(as usual, we should use a smooth integral cutoff at high frequencies).
In this way we get equations 
        \begin{eqnarray}
S_I(\omega )=S_0\coth \frac{eV}{2T}+2\omega e\Delta I \int_0^\infty 
{\cal A}(\tau )\, \sin (\omega \tau)\, d\tau 
        \nonumber\\
=S_0\coth \frac{eV}{2T}+2 e\Delta I \int_0^\infty 
\frac{d{\cal A}(\tau )}{d\tau}\, \cos (\omega \tau)\, d\tau , 
        \label{SviaA}\end{eqnarray}
which express $S_I(\omega )$ via  ${\cal A}(\tau )$ or 
${\dot {\cal A}}(\tau )$ [the last equation is obtained using 
integration by parts and taking into account ${\cal A}(0)=0$]. 

        To calculate ${\cal A}(\tau )$ (or ${\dot {\cal A}}$), we notice that 
the Bloch equations
(\ref{Bloch1})--(\ref{Bloch3}) couple the dynamics of ${\cal A}(\tau )$
with two more magnitudes
\begin{eqnarray}
 {\cal Y}(\tau )  \equiv && \sum_n n\, \mbox{Im} \rho_{12}^n (\tau ) \, ,
        \label{n-aver3}\\
 {\cal X}(\tau )  \equiv && \sum_n n\, \mbox{Re} \rho_{12}^n (\tau ) 
        \label{n-aver4} \, , 
\end{eqnarray}
        via equations 
\begin{eqnarray}
 \dot{\cal A} = && \frac{\Delta I}{2 e} - 4\, \frac{H}{\hbar}\, {\cal Y} 
+ 
                       \frac{I_0}{e}\, z \, ,
        \label{A-dot}\\
 \dot{\cal Y}  = && \frac{H}{\hbar}\, {\cal A} 
                  + \frac{\varepsilon}{\hbar}\, {\cal X} - \Gamma {\cal Y} 
    + b \, \mbox{Im} \rho_{12}  \, ,
        \label{Y-dot}\\
 \dot{\cal X}  = && - \frac{\varepsilon}{\hbar}\, {\cal Y} 
                        - \Gamma {\cal X}
        + b \, \mbox{Re} \rho_{12} \, , 
        \label{X-dot} 
\end{eqnarray}
        where $b= [(I_1^+ I_2^+)^{1/2}-(I_1^- I_2^-)^{1/2}]/e$. 
Because of the stationary initial conditions $z(\tau )=\rho_{12}(\tau )=0$,
so the equations are further simplified and become a closed system: 
\begin{eqnarray}
 \dot{\cal A} = && (\Delta I/2 e) - 4 (H/\hbar ) \, {\cal Y} ,
        \label{A-dot2}\\
 \dot{\cal Y}  = && (H/\hbar )\, {\cal A} 
                  + (\varepsilon /\hbar )\, {\cal X} - \Gamma {\cal Y} ,
        \label{Y-dot2}\\
 \dot{\cal X}  = && - (\varepsilon /\hbar )\, {\cal Y} 
                        - \Gamma {\cal X}.
        \label{X-dot2} 
\end{eqnarray}
        Solving these equations with the initial condition 
${\cal A}(0)={\cal X}(0)={\cal Y}(0)=0$, one can obtain ${\cal A}(\tau )$
and therefore $S_I(\omega )$.

        In the case of a symmetric qubit, $\varepsilon =0$, the evolution
of ${\cal X}$ is decoupled and one can find the analytical solution 
\begin{equation}
\frac{d{\cal A}(\tau )}{d\tau}  = \frac{\Delta I}{2 e}\, \exp{[-\Gamma \tau 
/2]}\, 
\left[ \cos {\tilde \Omega} \tau 
+ \frac{\Gamma}{2{\tilde \Omega}} \,
        \sin {\tilde \Omega} \tau \right] ,
\end{equation}
where ${\tilde\Omega}= \sqrt{\Omega^2-\Gamma^2/4}$. Substituting
this expression into Eq.\ (\ref{SviaA}) we finally obtain 
        \begin{equation}
S_I(\omega ) = S_0 \coth\frac{eV}{2T} + \frac{\Omega^2 (\Delta I)^2 \Gamma }
        {(\omega^2-\Omega^2)^2+\Gamma^2\omega^2} \, .
\label{result-sym} 
\end{equation}

        It is easy to check that at zero temperature this result coincides 
with the results of Refs.\ \cite{Kor-osc,Kor-Av,Av-book,Goan-osc}. 
Notice, however, that it does not assume weakly responding detector 
($|\Delta I|\ll I$) as in Refs.\ \cite{Kor-osc,Kor-Av,Av-book}. 
On the other hand, the derivation of Eq.\ (\ref{result-sym}) 
assumes low-transparency QPC as a detector, while a much broader class of 
linear detectors was considered in Refs.\ \cite{Kor-osc,Kor-Av,Av-book}.

\begin{figure}[t]
\centerline{
\epsfxsize=2.8in 
\epsfbox{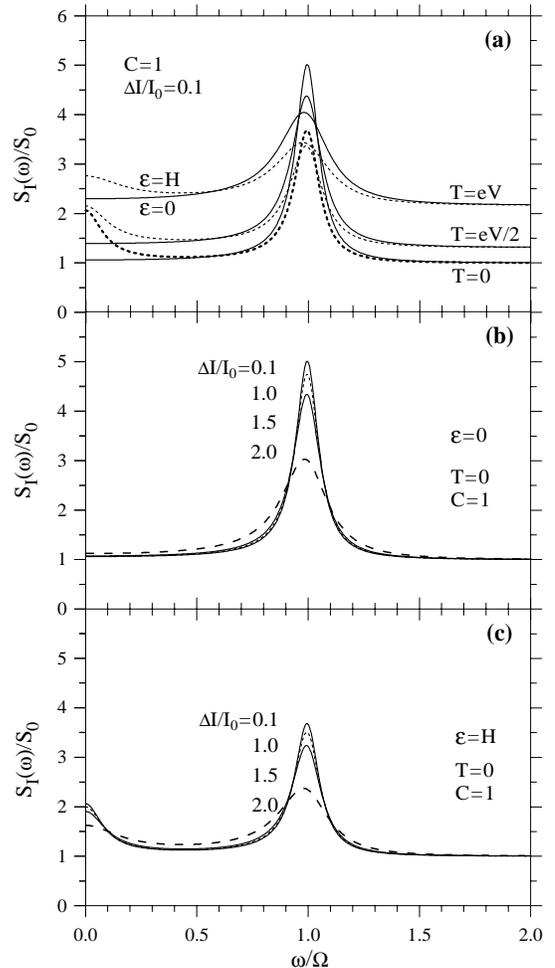}
} 
\vspace{0.4cm} 
\caption{(a): Spectral density  $S_I(\omega )$ of the detector current
in a weakly coupled (${\cal C}=1$) and weakly responding 
($\Delta I/I_0 =0.1$) regime at detector temperatures
$T=0$, $eV/2$, and $eV$ for a symmetric, $\varepsilon =0$ (solid lines), 
and asymmetric qubit with $\varepsilon =H$ (dotted lines). 
(b): $S_I(\omega )$ for a symmetric qubit and weakly coupled detector 
(${\cal C}=1$) at $T=0$ for several response ratios $\Delta I/I_0 =0.1$, 
1, 1.5, and 2. (c): the same as in (b) for an asymmetric qubit with 
$\varepsilon =H$. 
 }
\label{Fig3}\end{figure}

        With the temperature $T$ increase the noise pedestal $S_0\coth 
(eV/2T)$
increases while the spectral peak around $\Omega$ becomes lower and wider
[Fig.\ \ref{Fig3}(a)] 
because of $\Gamma$ increase [see Eq.\ (\ref{Gamma})]. The integral 
over the peak, 
        \begin{equation}
\int_0^\infty [S_I(\omega )-S_0\coth (eV/2T)] \, \frac{d\omega}{2\pi} = 
\frac{(\Delta I)^2}{4} \,  , 
\label{peak-integral}
\end{equation}
        does not depend on the temperature. (As will be seen later, this
formula remains valid for $\varepsilon \neq 0$ as well.)

        The peak-to-pedestal (``signal-to-noise'') ratio 
\begin{eqnarray}
        \frac{S_I(\Omega) -S_0\coth (eV/2T)}{S_0\coth (eV/2T)}= 
        \frac{(\Delta I)^2}{\Gamma S_0 \coth (eV/2T)}
        \nonumber \\
=\frac{4}{[\coth (eV/2T)]^2}\, \frac{\left(\sqrt{I_1}+\sqrt{I_2}\right)^2}
{2(I_1+I_2)} 
\end{eqnarray}
        has an upper bound equal to 4 and decreases with temperature 
as well as due to finite response ratio $|\Delta I|/I_0$. In particular,
in the case of very strong response, $|\Delta I|=2I_0$ ($I_2=0$), the
upper bound for the peak-to-pedestal ratio is 2 instead of 4 
[see Fig.\ \ref{Fig3}(b)]. Overall, the effect of the finite response 
on the spectral peak shape is similar to the effect of 
detector nonideality.\cite{Kor-osc}

        Let us emphasize that the derivation of Eq.\ (\ref{result-sym}) 
did not use any assumption about the magnitude of the coupling
${\cal C}\equiv \hbar (\Delta I)^2/S_0H$ between the qubit and detector,
so Eq.\ (\ref{result-sym}) remains valid even when the oscillations
are destroyed due to strong coupling (``quantum Zeno'' effect) 
and replaced by a telegraph noise. The analysis of the finite coupling effect
is completely similar to that of Refs.\ \cite{Kor-osc,Kor-Av}.
In particular, the quality factor of oscillations (disregarding the 
noise pedestal) is equal to 
        \begin{equation}
{\cal Q} =\frac{\Omega}{\Gamma}= 
\frac{8}{{\cal C}}\, \frac{\left(\sqrt{I_1}+\sqrt{I_2}\right)^2} 
{4I_0} \, \frac{1}{\coth (eV/2T)},
        \end{equation}
and the transition into overdamped regime occurs at ${\cal Q}<1/2$.

        For an asymmetric qubit, $\varepsilon \neq 0$, the analytical
solution of Eqs.\ (\ref{A-dot2})--(\ref{X-dot2}) is too lengthy,
so it is easier to use numerical calculations and then calculate the
Fourier transform (\ref{SviaA}) also numerically [Figs.\ \ref{Fig3}(a) and
\ref{Fig3}(c)].  
In the next Section we will show that the result is still 
equivalent to the results of Refs.\ \cite{Kor-osc,Kor-Av,Av-book,Goan-osc}
(within the common validity range).

        \section{Equivalence to the master equation approach} 

        Let us remind that the master equation approach\cite{Kor-osc,Kor-Av} 
assumes no correlation between the detector noise and qubit evolution
and treats the oscillating part of the current $I(t)$ as being proportional 
to the quantum operator ${\hat z}(t)$, so the spectral density $S_I(\omega )$ 
should be calculated as 
\begin{equation}
        S_I(\omega ) = S_{det} + \frac{(\Delta I)^2}{4}\, 
4 \int_0^{\infty} K_{\hat z}(\tau )\cos (\omega \tau )\, d\tau ,
\label{SviaKz}\end{equation}
where $S_{det}$ is the detector noise and $K_{\hat z}(\tau )\equiv
\langle {\hat z}(t+\tau ){\hat z}(t)\rangle$ is the correlation function
of ${\hat z}$. In the case of a weakly responding detector the detector noise
level does not depend on the qubit state and $S_{det}=S_0\coth (eV/2T)$.
The same formula remains valid in the case of moderate or strong
response because at high frequency $S_I(\infty )=2e(\langle I^+\rangle
+\langle I^-\rangle =2e\langle I\rangle \coth (eV/2T)$ and the average
current $\langle I\rangle$ remains to be equal to $I_0$.
As shown in Refs.\ \cite{Kor-osc,Kor-Av}, the correlation
function $K_{\hat z}$ is equal to the value of $z(\tau )=\rho_{11}(\tau )-
\rho_{22}(\tau )$ obtained from
the master equation (\ref{conv1})--(\ref{conv2}) with initial condition
$\rho_{11}(0)=1$, $\rho_{22}(0)=\rho_{12}(0)=0$.
(Actually, in those papers only the case of a weakly responding detector 
at zero temperature has been considered; however, the method can be easily 
generalized, since formally the only change in the master equation 
is a different $\Gamma$.) 

        Let us show that $S_I(\omega )$ calculated in this way coincides
with the result obtained from Eq.\ (\ref{SviaA}) at arbitrary qubit
asymmetry $\varepsilon$. For this purpose we introduce the new variable 
$a\equiv (2e/\Delta I){\dot {\cal A}}$ [so that we need to prove 
$a(\tau)=K_{\hat z}(\tau )$] and from Eqs.\ (\ref{A-dot2})--(\ref{X-dot2}) 
derive a new system of equations 
        \begin{eqnarray}
&& {\dot a}=-4(H/\hbar )\, y ,
        \\
&& {\dot y}=(\varepsilon /\hbar )\, x +(H/\hbar )\, a -\Gamma y ,
        \\
&& {\dot x}= -(\varepsilon /\hbar )\, y -\Gamma x, 
        \end{eqnarray}
where $y\equiv (2e/\Delta I)\, {\dot {\cal Y}}$ and $x\equiv (2e/\Delta I)\, 
{\dot {\cal X}}$. It is easy to see that these equations coincide with the 
master equations (\ref{conv1})--(\ref{conv2}) for $z$, $\mbox{Im}\rho_{12}$,
and $\mbox{Re}\rho_{12}$, respectively. Since ${\cal A}(0)={\cal X}(0)=
{\cal Y}(0)=0$, the initial conditions for new variables are
$a(0)=1$ and $x(0)=y(0)=0$, i.e.\ exactly the initial conditions for
$K_{\hat z}$ calculation. Therefore $a(\tau )=K_{\hat z}(\tau )$ and the
spectral density $S_I(\omega )$ calculated using the Bloch equations 
coincides with the result of the master equation approach.

	Hence, the analysis of $S_I(\omega )$ at finite qubit asymmetry 
$\varepsilon$ is completely similar to that of Refs.\ 
\cite{Kor-osc,Kor-Av,Av-book}. In particular, finite $\varepsilon$ 
leads to a decrease of the spectral peak around frequency $\Omega$ 
and origination of an extra peak around zero frequency (Fig.\ \ref{Fig3}), 
while the integral (\ref{peak-integral}) does not change [this is a 
consequence of $K_{\hat z}(0)=1$]. An analytical expression 
	\begin{eqnarray}
S_I(\omega) && = S_0 \coth \frac{eV}{2T}  + 
 \frac{\varepsilon^2 (\Delta I)^2 /(4H^2\Gamma )}
{1+(\omega\,\hbar^2\Omega^2/4H^2\Gamma )^2 } 
        \nonumber \\ 
&& +\,\frac{(\Delta I)^2 /[\Gamma (1+\varepsilon^2/2H^2)]} 
{1+\left[ (\omega -\Omega)\, 2/[\Gamma(1+\varepsilon^2/\hbar^2\Omega^2)]
\right] ^2} 
	\end{eqnarray}
can be obtained in the limit $\Gamma \ll \Omega$.

        Notice that for  both Bloch and master equation  approaches  
the  case  of  a  finite detector response 
does not formally differ from the case of a weakly responding detector 
(only the value of the ensemble decoherence rate $\Gamma$ changes).
For the Bayesian approach these two cases are significantly different,
so the generalization of the results\cite{Kor-osc} for $S_I(\omega )$
to a finite detector response (considered in the next Section) is not
trivial.

        \section{Generalization of the Bayesian results for $S_I(\omega )$} 

        The Bayesian results for $S_I(\omega )$ have been derived 
in Ref.\ \cite{Kor-osc} only for the case of a weakly responding detector,  
in which the detector current $I(t)$ can be considered as continuous and 
the qubit evolution is described by the equations\cite{Kor-99,Kor-rev}
        \begin{eqnarray}
\dot{\rho}_{11}= &&  -\dot{\rho}_{22}=
         \rho_{11}\rho_{22}\, \frac{2\Delta I}{S_0}\, [I(t)-I_0]
-2\,\frac{H}{\hbar}\,\mbox{Im}\,\rho_{12},
        \label{Bayes1}\\
 {\dot\rho}_{12}= && 
-( \rho_{11}-  \rho_{22})  \frac{\Delta I}{S_0} \,
[I(t)-I_0]\, \rho_{12} -\gamma \rho_{12} 
        \nonumber \\
&& {}  + \, {\rm i}\, \frac{\varepsilon}{\hbar }\,\rho_{12}+
        {\rm i} \, \frac{H}{\hbar } \, (\rho_{11}-\rho_{22})  ,
        \label{Bayes2}
        \end{eqnarray}
where $\gamma =\Gamma -(\Delta I)^2/4S_0$ is the qubit decoherence rate 
without ensemble averaging [for our model 
$\gamma /\Gamma =\cosh ^{-2}(eV/2T)$, so that $\gamma =0$ at $T=0$] 
and the statistics of $I(t)$ can be modeled as 
\begin{equation}
I(t) - I_0 =  [\rho_{11}(t) -\rho_{22}(t)] \, \Delta I/2 +\xi (t) ,
        \label{Bayes3}\end{equation}
where $\xi (t)$ is the white noise with the spectral density $S_\xi =S_0$.
Notice the significant difference in the meaning of $\rho_{ij}$ in the 
Bayesian equations and in the master equation since Eqs.\ 
(\ref{Bayes1})--(\ref{Bayes2}) describe individual qubit evolution without 
ensemble averaging. Also notice that the detector noise $\xi (t)$ is
now significantly correlated with the qubit evolution $\rho_{ij}(t)$. 

        To consider the case of a detector with finite response factor
$|\Delta I|/I_0$, we necessarily need to take into account individual 
tunnel events in the detector because $\Gamma$ becomes comparable
to $I_0/e$. Hence, the current is not continuous any more  
and we have to use generalized Bayesian equations\cite{Kor-rev} 
(which are essentially similar to the equations of the quantum jumps 
formalism\cite{Goan}). Then the qubit evolution during the time 
intervals between tunnel events in the detector is continuous and given by 
the small-time Bloch equations for $\rho_{ij}^{\, 0}$ with the restored 
normalization:
        \begin{eqnarray}
\dot{\rho}_{11}= &&  -\dot{\rho}_{22}=
        -\frac{\Delta I}{e}\, \coth (\frac{eV}{2T}) \, \rho_{11}\rho_{22} 
-2\,\frac{H}{\hbar}\,\mbox{Im}\,\rho_{12} ,
        \label{Bayes-gen1}\\
 {\dot\rho}_{12}= &&  
\frac{\Delta I}{2e} \, \coth (\frac{eV}{2T}) \, (\rho_{11}-\rho_{22})
\, \rho_{12} 
        \nonumber \\ 
&& + \, {\rm i}\, \frac{\varepsilon}{\hbar }\,\rho_{12}+
        {\rm i} \, \frac{H}{\hbar } \, (\rho_{11}-\rho_{22}) ,
        \end{eqnarray}
while each tunnel event in the detector (at time $t=t_k$) causes abrupt 
change (collapse) of the qubit state:
        \begin{eqnarray}
&&\rho_{11}(t_k+0) =\frac{I_1\rho_{11}(t_k-0)}
{I_1\rho_{11}(t_k-0)+I_2\rho_{22}(t_k-0)} \, , 
        \label{col1}\\
&&\rho_{22}(t_k+0)=1-\rho_{11}(t_k+0),
        \label{col2}\\ 
&&\frac{\rho_{12}(t_k+0)}{\rho_{12}(t_k-0)} = \left[ \frac{\rho_{11}(t_k+0)\,
\rho_{22}(t_k+0)}
{\rho_{11}(t_k-0)\,\rho_{22}(t_k-0)} \right] ^{1/2} .
        \label{col3}\end{eqnarray}
[Actually, in Eq.\ (\ref{col1}) instead of $I_i$ it is better to write $I_i^+$
if the tunneling is in the positive direction and $I_i^-$ if it is in
the negative direction; however, the corresponding temperature factors
cancel each other.] 
        It is interesting to note that the generalized Bayesian equations 
do not contain any decoherence term even at finite temperature.
This is because our model of the low-transparency QPC describes an ideal 
detector\cite{Kor-rev} and counting tunnel events in both directions
gives more information than measurement of only total current ($I^+-I^-$)
assumed in Eqs.\ (\ref{Bayes1})--(\ref{Bayes3}). 

        For the evolution simulation Eqs.\ (\ref{Bayes-gen1})--(\ref{col3})
should be complemented by the statistics of tunnel events in the detector.
This statistics is described by the (varying) rates $p^+$ and $p^-$
of tunneling events in the positive and negative directions, respectively:
        \begin{eqnarray}
p^+(t) = (I_1^+/e)\, \rho_{11}(t) + (I_2^+/e)\, \rho_{22}(t) ,
        \label{p+}\\
p^-(t) = (I_1^-/e)\, \rho_{11}(t) + (I_2^-/e)\, \rho_{22}(t) .
        \label{p-}\end{eqnarray}
It is important to notice that ensemble averaging of evolution equations 
(\ref{Bayes-gen1})--(\ref{col3}) over random moments of tunneling events
described by Eqs.\ (\ref{p+})--(\ref{p-}) leads\cite{Kor-rev} to the 
conventional master
equation (\ref{conv1})--(\ref{conv2}) with the ensemble decoherence rate
$\Gamma$ given by Eq.\ (\ref{Gamma}). 

        To calculate $S_I(\omega )$ we will use the method developed
in Ref.\ \cite{Kor-94} and write the current correlation function 
$K_I(\tau )\equiv \langle I(t+\tau )I(t)\rangle =K_I(-\tau )$ at 
$\tau \geq 0$ as 
        \begin{eqnarray}
K_I(\tau ) = && s\delta (\tau ) + \langle I^+\rangle \, e \, 
        [p^+(\tau | +)-p^-(\tau | +)]
        \nonumber \\
&&      -\, \langle I^-\rangle \,  e\, [p^+(\tau | -)-p^-(\tau | -)], 
        \label{K_I}\end{eqnarray}
where $s=S_I(\infty )/2$ determines the pedestal of $S_I(\omega )$ and
$p^\pm (\tau |\pm )$ is the average rate of tunneling in the positive ($p^+$)
or negative ($p^-$) direction at time $t+\tau$, for the condition 
that at time $t$ a tunneling in the positive ($|+$) or negative ($|-$)
direction has occurred. 

        The value of $s$ should be chosen in a way to provide the correct
value of $S_I(\infty )=S_0\coth (eV/2T)$ which can be calculated 
in the same manner as in the previous Section. For the calculation of 
$p^\pm (\tau |+)$ let us notice that as seen from Eq.\ (\ref{col1}), 
after the positive tunneling (at $\tau=0$) the value of 
$z=\rho_{11}-\rho_{22}$ is equal to 
        \begin{equation}
z(t+0|+) = \frac{I_1\rho_{11}-I_2\rho_{22}}{I_1\rho_{11}+I_2\rho_{22}}\, ,
        \label{z+}\end{equation}  
where $\rho_{ii}$ are taken before the tunneling. Averaging $z(t+0|+)$ over
the positive tunneling events [or, equivalently, over time with 
the weight factor $p^+(t)$, which  
is proportional to the denominator of Eq.\ (\ref{z+})], 
we get $\langle z(t+0|+)\rangle = \langle I_1\rho_{11}-I_2\rho_{22}\rangle
/\langle I\rangle$, expressed via simple averaging over time. Since
$\langle\rho_{11}\rangle = \langle \rho_{22}\rangle =1/2$, the expression
can be further simplified: 
$\langle z(t+0|+)\rangle = \Delta I/2 \langle I\rangle$. 
Similar calculation shows that 
$\langle \rho_{12}(t+0|+)\rangle = \langle \rho_{12}\rangle (I_1I_2)^{1/2}
 /\langle I\rangle = 0$. 

      It is sufficient to know $\langle z(t+0|+)\rangle$ and
 $\langle \rho_{12}(t+0|+)\rangle$ to calculate
$p^\pm (\tau |+)$ because of the linearity of the averaged evolution 
equations (\ref{conv1})--(\ref{conv2}) in terms of $z$ and $\rho_{12}$
and linearity of Eqs.\ (\ref{p+})--(\ref{p-}). 
Using Eqs.\ (\ref{conv1})--(\ref{conv2}) with averaged initial conditions
at $\tau =0$, we can show 
        \begin{equation}
p^\pm(\tau |+) -\frac{\langle I^\pm\rangle }{e} =  
\frac{I_1^{\pm}-I_2^\pm }{2e} \, 
\frac{\Delta I}{2\langle I\rangle } \, z(\tau ), 
        \end{equation}
where $z(\tau )$ is calculated from Eqs.\ (\ref{conv1})--(\ref{conv2})
starting from initial condition $\rho_{11}=1$, $\rho_{22}=\rho_{12}=0$.
        Finally noticing that the expressions do not depend on
the direction of tunneling at $\tau =0$ and combining the terms in
Eq.\ (\ref{K_I}) we obtain 
\begin{equation}
K_I(\tau >0) = \langle I\rangle ^2 +\frac{(\Delta I)^2}{4}\, z(\tau ) 
\label{K_I-2}\end{equation}
with the same $z(\tau )$ as above. 
The constant term $\langle I\rangle ^2$ does not contribute to the Fourier
transform 
$S_I(\omega )= 2 \int_{-\infty}^\infty K_I(\tau )\cos (\omega \tau) \, 
d\tau$  (formally it leads to a $\delta$-function at $\omega =0$),
while the second term of Eq.\ (\ref{K_I-2}) gives the same contribution 
as the second term of Eq.\ (\ref{SviaKz}). Consequently, the calculation
of $S_I(\omega )$ using the generalized Bayesian approach leads to the
same result as the master equation approach.

        \section{Conclusion}

        The main result of this paper is the development of a method 
of $S_I(\omega )$ calculation for a detector measuring 
quantum coherent oscillations of a qubit, based on the microscopic
Bloch equations,\cite{Gurvitz} which couple the qubit and detector degrees 
of freedom. As the detector we assumed a low-transparency QPC 
(tunnel junction). 
        We have shown that $S_I(\omega )$ calculated in this way formally 
coincides with the results obtained previously 
\cite{Kor-osc,Kor-Av,Av-book,Goan-osc} by the Bayesian, master equation, 
and quantum jumps methods, though in this paper we have considered a wider 
validity range. In particular, 
our formalism takes into account finite detector temperature $T$ and 
assumes arbitrary detector response $|\Delta I|/I_0$.

        Besides that, we have generalized the Bayesian method of 
$S_I(\omega )$ 
calculation to the case of arbitrary detector response and temperature. 
(The generalization of the master equation method is formally trivial.) 
We have proven that the results of all three methods still coincide 
in such generalized case.

        The model we have considered describes essentially an ideal detector.
The detector nonideality can be phenomenologically taken into account 
by introducing an extra dephasing term into the Bloch, master, 
and Bayesian equations, containing $\dot{\rho}_{12}$. 
This will lead to an increase of the ensemble decoherence 
rate $\Gamma $ and, therefore, to a wider and lower peak of $S_I(\omega )$,
corresponding to the qubit oscillations.\cite{Kor-osc,Kor-Av,Av-book,Goan-osc}
Similar procedure can be done to take into account a qubit-controlled change 
of the detector tunneling phase\cite{Kor-Av,Av-book,Goan-osc} 
[then $\Gamma =( 2I_0- \sqrt{4I_0^2-(\Delta I/\cos \theta)^2}\, ) 
(2e)^{-1} \coth (eV/2T)$ where $\theta =\arg (M^*\Delta M)$] 
even though such detector is still ideal in the 
generalized sense.\cite{Kor-rev,Av-book,Goan-osc} 
The effect of a weak extra coupling between the qubit and 
a passive finite-temperature environment (with temperature different from 
$T$) can be taken into account in a way similar to Refs.\ 
\cite{Kor-osc,Kor-Av,Av-book}. 
We did not consider these effects in the present paper because their
treatment is exactly the same as in previous papers. 

        An experimental measurement of $S_I(\omega )$ and verification
of the upper bound ($\leq 4S_0$) for the spectral peak corresponding to qubit
oscillations seems to be the easiest experiment related to a continuous 
monitoring of a nontrivial single qubit evolution. This makes it 
preferable for sooner realization in comparison with more interesting 
but more difficult proposed experiments\cite{Kor-99,Ruskov,Kor-Bell} 
on monitoring and continuous collapse of a solid-state qubit.

\vspace{0.3cm} 
        The authors thank L. Fedichkin for useful discussions. 
The work was supported by NSA and ARDA under ARO grant DAAD19-01-1-0491.


\begin{references}
\bibitem{eV} In particular, we should always assume $eV \gg \hbar\Omega$. 
       	To consider individual electron tunneling in the detector, we also
	need $eV/\hbar \gg I/e$ that translates into a condition of 
	sufficiently large resistance: $R\gg \hbar /e^2$ (even for a 
	multimode QPC). However, in the 
	weakly responding regime the second condition can be relaxed to 
	$eV/\hbar \gg (\Delta I)^2/S_0$ that translates into $R \gg 
	(\Delta I/I_0)^2 \hbar /e^2$ (the transparency of each mode is still
	small).
\end{references}
\end{document}